\documentclass[sigconf]{acmart}
\AtBeginDocument{%
  }

\setcopyright{acmlicensed}
\copyrightyear{2026}
\acmYear{2026}
\acmDOI{XXX.YYY}
\acmConference[]{XXX}{YYY}{ZZZ}
\usepackage{array,tabularx}

\usepackage{dblfloatfix}
\usepackage{graphicx}
\usepackage{multirow}
\usepackage{multicol}
\usepackage{comment}
\usepackage{xspace}
\usepackage{makecell}
\setcellgapes{3.5pt}

\def\framework{\textbf{\textsc{MUSBe}}\xspace}
\def\method{\textbf{\texttt{SpecDoc}}\xspace}

\begin{document}

\title{Modeling User-System Behavior for \\Training-free Building of Private Domain Conversational Agents}

\author{Won Ik Cho, Woonghee Han, Kyung Seo Ki, Young Min Kim}
\affiliation{%
  \institution{AI Center, Samsung Electronics}
  \city{Suwon}
  \country{Korea}
}

\renewcommand{\shortauthors}{Cho et al.}

\begin{abstract}
  The rise of agentic systems that combine orchestration, tool use, and conversational capabilities, has been more visible by the recent advent of large language models (LLMs). While open-domain frameworks exist, applying them in private domains remains difficult due to heterogeneous tool formats, domain-specific jargon, restricted accessibility of APIs, and complex governance. Conventional solutions, such as fine-tuning on synthetic dialogue data, are burdensome and brittle under domain shifts, and risk degrading general performance. In this light, we introduce a framework for constructing private-domain tool-calling multi-agent systems that does not necessitate data generation and model tuning, by modeling user and system behaviors via tool-use documentation. With a simplified composition that assumes an orchestrator, a tool-calling agent, and a general chat agent, we show that the proposed approach enables scalable adaptation to private tools, which had been previously achieved by accompanying continual retraining. Our framework supports practical use cases, including lightweight deployment of in-house multi-agent systems, providing a sustainable method for aligning agent behavior with domain expertise in private, enterprise-level conversational ecosystems.
\end{abstract}





\maketitle

\section{Introduction}

The development of agentic AI systems that can orchestrate multiple tools and engage in interactive conversations, has been more visible by the advent of large language models (LLMs) that can role as agents. These AI agents can dynamically decide which external tool or API to invoke in order to fulfill a user’s request, leveraging the reasoning and language understanding capabilities of LLMs. Several techniques and frameworks have been introduced to facilitate such multi-tool calling behavior \cite{schick2023toolformer,shen2023hugginggpt,qin2024toolllm}. 
Beyond research prototypes, practical tool-use platforms have emerged. The model context protocol (MCP), for example, is a standardized interface that allows an AI agent to discover and invoke a collection of tools through a unified client–server API \cite{MCP2023}. Such tool orchestration toolkits, along with agent frameworks like Agno (an open-source library for building multi-agent systems), are freely available for developers \cite{agno2025docs}. These resources 
enable building multi-tool agentic systems in general domains on local infrastructure, assuming sufficient computational resources are available for running the LLM and associated tools.

However, implementing agentic AI in private and domain-specific settings, especially within firms of specialized industry, remains highly challenging. Public agentic AI frameworks and pre-trained models often assume open-domain knowledge and standard tool APIs; when applied to a proprietary domain, they typically require additional adaptation or fine-tuning \cite{gururangan2020dont}. Enterprises usually possess internally-developed tools (e.g. custom databases, analytics APIs) with heterogeneous formats and I/O conventions, which are not standardized across the organization. Integrating such unstandardized internal tools can confuse an LLM-based agent, since each tool may expect inputs/outputs in a different form that the model was never trained on. The difficulties include: (1) non-unified calling conventions and data formats for tools, arising from diverse development practices, which are addressed by frameworks like MCP that attempt to impose a standard interface but usually unprepared in-house; (2) domain-specific or proprietary jargon that the backbone LLM may not fully understand \cite{ling2023domain} limiting the model’s ability to leverage expert knowledge without additional training; (3) restricted accessibility of the tools, since they are deployed in a closed environment with strict authentication and cannot be freely accessed via commercial APIs (unlike tools on the open web) \cite{jeong2023study}; and (4) complex governance requirements, as multiple stakeholders (e.g. compliance officers, IT administrators, domain experts) impose constraints on how data and tools can be used, which is difficult to encode directly into an agent’s policy.

To tackle these challenges, prior approaches have often relied on creating synthetic dialogue data to fine-tune LLMs for the target domain or toolset \cite{tang2023toolalpaca,wu2024sealtools,shim2025tooldial}. In such solutions, developers and domain experts imagine example conversations where a user interacts with the system, including invocations of the private tools; large quantities of these synthetic dialogues are then generated (sometimes via self-play or in-house sourcing) to train or evaluate the agents. While fine-tuning can indeed imprint domain-specific behavior on an already powerful LLM, it carries the risk of degrading the model’s general capabilities outside that domain due to catastrophic forgetting \cite{luo2025empirical}. Recent studies have observed that even fine-tuning on relatively benign, domain-specific data can partially erode the original model’s aligned behaviors and performance on other tasks \cite{kumar2022fine,frasers2025fine}. Mitigating this would require sophisticated continual learning techniques—e.g. intermixing diverse types of training data (reasoning tasks, generic instructions, multilingual data, etc.) during domain adaptation to preserve broad knowledge \cite{ramponi2020survey,wei2022flan}. Although such comprehensive retraining is ideal in theory, it is often impractical in enterprise settings due to limitations in available in-domain data, lack of computing infrastructure for large-scale training, and uncertainty in hyperparameter tuning for stability.

LLMs (either vanilla pre-trained or fine-tuned versions) are expected to serve as stand-alone conversational agents for many scenarios, but there is growing recognition that a multi-agent architecture can be more effective for complex conversational systems \cite{guo2024llmagent}. Relying on a single monolithic agent to handle all aspects of dialogue—understanding user intent, retrieving knowledge, performing actions, and generating responses—may be burdensome even for state-of-the-art models. In fact, classical dialogue systems were explicitly modular, with separate components for language understanding, state tracking, action policy, and language generation \cite{gao2019neural}. Expecting one LLM to internally learn and execute all these functions increases the risk of errors, especially in domain-specific contexts where understanding implicit intents and calling relevant tools are both required \cite{zhao2022unids}. Multi-agent frameworks alleviate the load on each model by assigning different roles to specialized agents, allowing each to focus on what it does best \cite{dorri2018multi,chen2023agentverse,guo2024llmagent}. Indeed, the remarkable capabilities of the latest LLMs help us compose systems of multiple LLM agents, rather than over-burdening a single model.

In this light, we describe a practical building method for multi-agent conversational framework tailored to private-domain tool use. Our framework, conceptually suggested assuming a minimal system with general chat and tool-calling ability, achieves tool integration without any model fine-tuning; instead, we inject domain knowledge and usage conventions through structured behavior modeling and extensive documentation. In summary, we propose (1) a \textbf{conceptual framework} \framework (\textbf{M}odeling \textbf{U}ser-\textbf{S}ystem \textbf{Be}havior, pronounced \textit{musbi}) for building private-domain multi-agent systems, defining the key stakeholders and components involved; and (2) a \textbf{training-free method of using} \method for aligning an LLM-based agent to domain-specific tools and jargon by means of behavior specification and documentation, instead of labor-intensive data (re-)generation. We describe a concrete implementation of this framework and demonstrate the principle use case --  the development and deployment of the private, domain-specific multi-agent system -- compared with traditional model-tuning approaches, accompanying operational observations and lesson-learned from the deployment experience.

\section{Background}

\subsection{Agents and Multi-Agent Systems in LLM Era}
In AI terms, an agent is an entity that perceives its environment and makes decisions or takes actions autonomously to achieve goals \cite{ferber1999multi,russell2010ai,qu2025comprehensive}. Traditional agent research often emphasizes properties like reactivity, proactiveness, and social ability of agents, and in multi-agent systems (MAS) these agents interact or cooperate within an environment. Recently, there has been debate on how the emergence of LLM-based agents fits into classical MAS definitions \cite{he2025llm}. Many so-called ``agentic’’ LLM systems lack hallmarks of true MAS such as coordination, cooperation, and negotiation \cite{buadicua2025contemporary}, instead being closer to single-agent prompts that sequentially invoke tools or other LLMs \cite{guo2024llmagent}. Here we adopt a practical view: we consider an LLM-based agent to be an LLM or a module built on one that is situated in a loop of observation and action (via language), possibly among other agents, to fulfill certain tasks.

In the context of LLM applications, a multi-agent system typically refers to a setup in which multiple LLMs (or LLM-powered components) specialize in different sub-tasks and communicate or cooperate to solve a complex problem \cite{chen2023agentverse}. For example, one agent might function as a planner or manager, breaking a user query into sub-tasks, and assigning these to other agents specialized in tools or knowledge domains. Such orchestrator–subagent architectures have been explored as a way to handle complex multi-step tasks more effectively than a single prompt model \cite{hong2023metagpt}. This paradigm differs from classic modular NLP pipelines in that the ``modules’’ here are often homogeneous (all are LLMs) and communicate in natural language, but it shares the spirit of divide-and-conquer by skill specialization. 
Overall, when carefully designed, multi-agent architectures marry the robustness of modular systems by each part doing a limited job well, with the adaptability of LLMs -- using prompts to define roles rather than hard-coding logic \cite{shankar2024pipeline}.

A common multi-agent pattern in recent LLM systems is the controller or router + tool-using agents model. This entails one agent that receives user input and decides whether to invoke a tool or to respond directly. If a tool is needed, the input (or a processed form of it) is delegated to a sub-agent that is adept at using tools or APIs to act on the query. This agent then returns a tool output which the first agent may wrap into a final answer for the user. This design is seen in systems like HuggingGPT (ChatGPT orchestrating domain-specific models) and MetaGPT (an LLM acting as a “Project Manager” coordinating other LLM agents in roles like Engineer, Architect, etc.) \cite{shen2023hugginggpt, hong2023metagpt}. Compared to traditional pipeline dialog systems, which also had distinct components (natural language understanding and generation, dialogue state tracking, and policy) \cite{gao2019neural}, the modern orchestrator–subagent approach still leverages specialization but often uses the same underlying model for multiple agents, differing only by prompt or a lightweight fine-tuning. 

\subsection{LLM Adaptation: Fine-Tuning vs. Prompting}

LLMs can be adapted to new tasks or domains either through additional training of their parameters (fine-tuning) or through clever prompt design and in-context learning. In the pre-LLM era of ``pre-trained language models’’ (PLMs) like BERT \cite{devlin2019bert}, fine-tuning on task-specific data was the dominant approach \cite{gururangan2020dont}. Numerous strategies were explored, from full-model fine-tuning to more parameter-efficient methods such as adding adapter layers \cite{houlsby2019adapter} or low-rank parameter updates \cite{hu2022lora}. As model sizes grew into the tens of billions of parameters, these parameter-efficient fine-tuning (PEFT) techniques, as well as continual pre-training on domain data, became vital to adapt models without exorbitant costs. 

Fine-tuning an LLM on dialogues or instructions tailored to a private domain is an intuitive solution for building a conversational agent for that domain. 
However, as discussed, this comes with several practical difficulties when the domain is highly specialized. First, constructing a sufficiently large and representative fine-tuning dataset is often the hardest bottleneck \cite{ramponi2020survey}. If the domain data cannot be released or even centralized due to privacy, data augmentation via external APIs or crowdsourcing is infeasible \cite{kulkarni2023llms}. Developers are then left to simulate user interactions, an error-prone and labor-intensive process. Moreover, even if an LLM is fine-tuned to call a certain set of tools correctly, that training may become obsolete when the tools are updated or new tools are added, necessitating frequent re-training of the model. This is clearly problematic for rapidly evolving private software environments.
Given these issues, recent trends have shifted toward leveraging the raw power of LLMs through prompting, without additional training, for custom applications. Techniques such as providing a descriptive system prompt (persona) and a few demonstrations in context can coax a pre-trained model to behave in domain-specific ways \cite{liu2023pretrainprompt,min2022rethinking}. 

This philosophy extends naturally to multi-agent setups. Rather than training separate models for each agent’s role, one can use the same base LLM with different role prompts or contexts for each agent \cite{park2023generative}.
Early studies indicate that such prompt-based multi-agent systems can achieve strong performance if the underlying model is strong, and they are far easier to adjust or maintain (one can update a prompt or add a new tool agent without any gradient updates) \cite{shen2023hugginggpt}. There is no established ``best practice” yet on when a multi-agent architecture is warranted or how to optimally prompt each agent; the design remains task-dependent. Nonetheless, given the continuous improvement in LLM capabilities and context lengths, it is a promising direction for domain-specific conversational AI: we can compose a few instances of a powerful general model, each guided by domain-informed instructions, to achieve specialized functionality without additional training.

\subsection{Challenges in Specialized Domains}

Adapting AI agents to specialized domains (such as manufacturing, healthcare, finance, or law) has long been an area of interest in the ML community, typically under the umbrella of domain adaptation or transfer learning \cite{ramponi2020survey}. In the era of deep learning and PLMs, the common recipe was to start from a model pre-trained on general data (e.g. Wikipedia, Books) and then continue training it on domain-specific data to create a domain-specialist model such as BioBERT for biomedical text mining \cite{lee2019biobert} and SciBERT for scientific papers \cite{beltagy2019scibert} etc. This approach has seen success in improving performance on domain-specific NLP tasks by injecting domain knowledge into the model’s parameters. However, even these specialist models assume the availability of sizable domain corpora (BioBERT with PubMed articles, SciBERT with Semantic Scholar papers). Moreover, for truly private domains like a specific enterprise’s internal data or tool, such large-scale high-quality text might not be obtainable or sharable.

The advent of LLM-based agents, which not only produce text but take actions, introduces further complications in under-studied domains. Some domains with plenty of textual data have attracted significant LLM research, but others like manufacturing or enterprise IT management have seen comparatively little public work, despite their economic importance. One reason is that organizations in these sectors are extremely cautious about data privacy and security \cite{jeong2023study,kulkarni2023llms}. They cannot simply leverage public APIs or services that involve sending proprietary data off-site. Any conversational assistant or agent must be built and deployed in-house, operating only on local data and behind firewalls. This limits the ability to use cloud-based foundation models and also complicates the data collection for training. 
In short, many vertical domains face a cold-start problem: they need conversational AI, but they have neither existing dialogue data nor permission to use external AI services to create such data.

Another challenge is that internal tools and software in these domains tend to be moving targets. In manufacturing or enterprise software, for example, APIs and databases are updated frequently and at most times without consideration of LLM-based calling. A fine-tuned model that was trained on last quarter’s tool specifications may become inconsistent with the current system behavior. Traditional dataset collection and model retraining would lag behind the actual system changes. Thus, a static dataset-to-training pipeline is ill-suited to these scenarios. What is needed instead is a more dynamic and maintainable approach, where updating the agent’s knowledge of the tools does not require starting a new data collection or fine-tuning cycle from scratch.

These observations motivate our work. We aim to develop a framework for human-in-the-loop behavior modeling that allows system developers and domain experts to inject the necessary domain knowledge and conventions directly into a conversational agents, without requiring massive dialogue datasets. By treating the problem as one of documentation and specification rather than making the model memorize, we seek to create private-domain conversational agents that are easier to build and update over time.

\section{Multi-Agent System for Private Domains}

We begin by outlining the conceptual design of our multi-agent conversational system $S$ intended for a private domain $D$. The domain $D$ is characterized by both domain-specific knowledge which might include specialized terminology and workflows that generic LLMs don’t fully grasp, and organization-specific privacy constraints that preclude using external API calls or sharing data publicly. The system $S$ is envisioned as a chat-based assistant used internally by people in domain $D$. For generality, one can imagine $D$ to be a large enterprise’s manufacturing division, though the design would apply similarly to, say, a proprietary financial analytics platform or a medical information system accessible within a hospital. What distinguishes $D$ from open domains is that the relevant tools, data, and jargon are not part of the common web knowledge; unlike broad domains such as general medical or financial knowledge, a private domain’s knowledge is often local and not standardized across different organizations.

There are several stakeholder groups involved in building and using $S$: (a) the \textbf{system developers} which can be further divided into roles like data manager, model engineer, and infrastructure engineer, (b) the \textbf{tool providers} – the engineers or teams who maintain the internal APIs and databases that $S$ will call, (c) the \textbf{domain experts }– experienced users or subject matter experts in $D$ (who might not be familiar with AI, but know what $S$ needs to do), and (d) the \textbf{end users} of $S$ within the organization. 

Traditional approaches to creating a domain-specific system would require intensive collaboration between these stakeholders to produce training data \cite{seymoens2018methodology,hoogland2021exploring,kumar2022data}. A data manager would ask domain experts for example scenarios, then create synthetic dialogues covering those, and finally a model engineer would fine-tune the LLMs on this data \cite{choi2023dmops}. However, such procedure can be quite onerous if the tools have never been accessed via natural language before, that there are no real conversation logs to learn from – every scenario must be imagined by the creators. The domain experts play a crucial role in seeding this process with realistic use cases \cite{rastogi2020sgd,oakes2024building,song2025injecting}, which the data team then expands into a training corpus through paraphrasing, augmentation, etc. Essentially, the domain knowledge is injected via these manually scripted dialogues, hoping that the model will internalize the patterns.

\subsection{Multi-Agent System: The Composition}

In our multi-agent framework, we propose a more direct injection of domain knowledge through modeling the behavior of end users instead of collecting thousands of example dialogues. For simplification, we assume the multi-agent system $S$ that includes at least the following LLM-based agents: an orchestrator, a tool-calling agent (TCA), and a general chat agent (GCA). This minimal set is inspired by classic task-oriented dialogue systems which separate task-specific and chitchat capabilities \cite{zhao2022unids}, as well as by recent multi-agent research that often uses a task router plus task executors \cite{shen2023hugginggpt}. We deliberately keep the agent types few and broad: the TCA is responsible for interacting with domain tools and retrieving factual answers, while the GCA handles open-domain conversation or off-topic user input, including applying safety or refusal policies if needed. The orchestrator’s job is to examine each user query along with conversation context and decide which agent should handle it. This division of labor aligns with the intuition that not all queries require an action or tool call; by routing those to GCA, we avoid overloading the TCA with unrelated dialogue. Conversely, when a query does pertain to the domain’s tools or data, the orchestrator should invoke the TCA.

Notably, we do not assume any of these agents are fine-tuned to the domain. They could all be instances of the same base LLM loaded with different system prompts, and if resources allow, one could choose more suitable one for each agent. But in our implementation, we focus on using a single base model for simplicity.

The \textbf{Orchestrator} is analogous to a dialogue manager. It does not produce end-user answers by itself; rather, it decides which agent should respond at each turn and how to route information between agents. A typical orchestrator prompt might say: \textit{``You are the orchestrator of a multi-agent system for domain $D$. Analyze the user’s request and decide whether it should be handled by the tool-calling agent or the general chat agent. Do not answer the user directly. Your output should be either: TOOL\_CALLING\_AGENT (and then you will receive the TCA’s answer to forward) or GENERAL\_CHAT\_AGENT (and then forward the GCA’s answer).’’} We also supply the orchestrator with descriptions of each agent’s capabilities and some heuristic rules for borderline cases; for instance, if the user’s query contains domain-specific jargon or mentions a tool name, route to Tool Agent, unless it’s a general question about how to use the system, etc. The orchestrator thus needs some awareness of domain terminology to classify intents correctly. 

The \textbf{Tool-Calling Agent} is the workhorse that interfaces with the domain’s tools and data sources. Its prompt might begin: \textit{``You are the Tool Agent in a multi-agent system for $D$. Your job is to decide which API or tool (from the provided list) can answer the user’s query, call the tool with appropriate parameters, and then present the result.’’} We provide it with guidelines on how to interpret colloquial or jargon-laden user requests into proper API calls. Importantly, it is given access to documentation of the domain’s tools (in a form we will elaborate in Section~\ref{sec:framework}) so that it can match user intents to specific API functions and fill in required parameters. It also has rules on formatting the calls and on what to do if multiple steps are needed. In essence, the TCA’s behavior is largely dictated by how comprehensively and clearly we feed it the tool specification and usage conventions.

The \textbf{General Chat Agent} is a fallback that covers anything outside the domain or any casual conversation. Its prompt might be the simplest: \textit{``You are the general chat agent. You can handle general questions or small talk. If the user asks about domain-specific tools, you should refuse and suggest they ask the tool-calling agent.’’} Additionally, in producing direct answers to the user for open-domain questions, one can include any necessary safety instructions, e.g. not revealing confidential info, following enterprise AI guidelines; in our framework, the GCA is not expected to provide domain expertise which is the TCA’s role. It primarily ensures that the user experience is smooth and safe.

\subsection{Justification of the Structural Setting}

We decided to adopt the orchestrator + TCA + GCA hierarchy because it directly corresponds with the desired capability of $S$ in the traditional approach we had taken, so that the comparison between two approaches can be conducted more fairly. In the very first attempt using synthetic dialogues, we first created tool-calling dialogues with conventional \textit{user}, \textit{assistant}, and \textit{json}-based tool-argument pairs, and trained LLMs so that they react to specific types of user queries. However, LLMs displayed the behavior overfitted to the synthetic data and forgot how to interact with the user. Thus, we secured the ability to chit-chat by augmenting greetings and small talks before and after the tool calling part, so as to achieve the multi-turn sustainability of the model. 

Our minimal configuration mirrors a typical scenario: sometimes the user just wants a straightforward answer or a chat, other times they need some data or action done via internal tools. We assume that question-answering (QA), one of the basic capabilities expected for agentic systems, is also one form of tool-calling if it necessitates private and domain-specific knowledge, and otherwise handled by GCA. One could extend the framework with more agents if needed, but we focus on this minimum viable multi-agent system for clarity.

Crucially, even with these two functional agents, domain knowledge injection is needed in at least two points: (1) the orchestrator needs to recognize domain-specific cues in the user’s request to hand it to the TCA when appropriate, and (2) the TCA needs to correctly interpret domain-specific language and map it to tool usage. In conventional systems, both the orchestration and the API-calling policy had been learned data-intensively \cite{henderson2014dstc,gao2019neural,shim2025tooldial}. 
However, to avoid entity overfitting \cite{vm2024fine} and catastrophic forgetting \cite{luo2025empirical}, we supply this knowledge explicitly via instructions or structured information, avoiding the generation and training of synthetic data.

\section{Behavior Modeling and Its Implementation}
\label{sec:framework}

Our proposed solution for building a private-domain multi-agent system without training is a process we call \framework, inspired by user modeling strategies \cite{tan2023user,raggioli2025comparing,shah2025using}. This involves creating explicit structured representations of the agent behaviors and domain knowledge, instead of implicit learning. The core of this approach is a comprehensive Specification Document (\method) inspired by the concept suggested in structured software process modeling \cite{franch1998structured}, that details all relevant aspects of the tools (APIs) available and the conventions of their use in the domain. Think of the document as a hybrid of an API reference manual, a domain glossary, and dialogue guidelines. It is written in natural language with structured sections, targeted not at human end-users but at the LLM agents themselves. By providing this document to the agents via prompting, we aim to supply the missing domain-specific knowledge directly.

To illustrate, imagine a manufacturing domain with an internal API \textit{GetMachineStatus(machine\_id)} that returns the status of a machine by ID. An official API document might say: \textit{\{Function: GetMachineStatus; Input: machine\_id (string); Output: status (string, one of {Running, Stopped, Maintenance}); Description: returns the current status of the machine.\}} Our Spec Document entry for this tool would expand this to include: what users typically say when they want a machine’s status (\textit{``Is machine 7 up right now?’’} or \textit{``check if machine 7 is down’’}), how to map those utterances to the \textit{machine\_id} parameter, what the output means (if \textit{status} = Maintenance, perhaps the user would say \textit{``under maintenance”}), any related follow-up actions (like if a machine is down, the next query might be \textit{``who is the technician on duty?”} which might necessitate another API). Essentially, we model the behavior of both users and the system around this tool, which incorporate the domain-specific convention that might never be pretrained in publicly available AI models.

\method thus serves as a data abstraction layer. Instead of collecting a thousand example dialogues exhibiting this behavior, we write it down declaratively. Of course, this is a human-intensive process too, but it offers better scalability: adding a new tool means writing its spec entry rather than collecting new dialogues for it, which is more straightforward for developers and tool providers. It also inherently supports updates: if a tool’s behavior changes, one can update the specification text accordingly and the agents with updated prompts will adjust.

Below is a general template of what \method for a tool (here, \textit{GetMachineStatus}) or tool group might contain:

\begin{quote}\small
 
\textbf{1) Purpose}: Brief description of the tool’s function and when it should be called.

\textbf{2) Provider / Data Source}: Who provides this tool, where does its source data come from, any terms of usage or access conditions.

\textbf{3) Inputs (Arguments)}: List each input parameter, its type, meaning, and any domain-specific terms or aliases that users might use to refer to it. For example: \textit{machine\_id}: the unique identifier of a machine. Users might refer to this as "machine number", "ID tag", or by a name (see mapping rules).

\textbf{4) Outputs}: Describe the output structure or fields returned. Include how these might be expressed in natural language. For instance: status: operational status of the machine. Possible values are \textit{``Running", ``Stopped", ``Maintenance"}. Users might say \textit{``up", ``down"}, or \textit{``ongoing"}.

\textbf{5) Slot-Filling Phrases}: Common phrases or jargon for providing each argument. E.g. for \textit{machine\_id}, users often say \textit{``machine 247"} or just \textit{``m247"} to mean machine\_id=``247". The tool agent should extract the number accordingly.

\textbf{6) Output Post-processing}: If the raw output needs formatting or triggers a follow-up action. E.g. if status=Stopped, the system should suggest checking error logs (another tool).

\textbf{7) Visualization / UI Action}: (If applicable) How the output should be displayed or whether the system should produce a chart, etc. Describe any special handling in the front-end.

\textbf{8) Default Behaviors}: What the agent should do if some inputs are missing or if output is empty (ask the user again or set default parameters). 

\textbf{9) Related Tools}: Tools that are often used before or after this tool. For example: After getting machine status, if status is Down, consider calling GetDowntimeReason. List any natural language cues that indicate a transition.

\textbf{10) Contextual Usage}: Any special context or conditions. E.g., 1) This tool should only be used during working hours queries. 2) If the user asks about "all machines status", a different tool is needed instead. (...)
 \end{quote}

By organizing the domain knowledge in this manner, we aim to cover everything a dialogue model would need to know to handle the tool correctly. The guiding principle is to anticipate what a fine-tuned model would have ``learned” from dialogues and to provide that information explicitly. The authors of \method, likely a collaboration between tool developers and the AI team’s data experts, must keep in mind that their text will be consumed by an LLM, not a human end-user. Thus, the writing should be precise, unambiguous, and as LLM-comprehensible as possible – for instance, avoiding overly complex sentence structures, and clearly delineating examples.

The authors of the document should also remember that, while it stands in place of training data and will be parsed by the LLM at runtime, it still incorporates some ambiguity as a nature of natural language. Therefore, consistency and clarity are paramount. If two tools have a parameter with the same name but different meaning, that should be explicitly highlighted to avoid confusion. Ideally, naming conventions for functions or arguments could be aligned to reduce ambiguity – this might involve liaising with tool developers to perhaps alias parameters similarly across tools where possible. But where that’s not possible, \method should point it out so the agent doesn’t incorrectly assume.

In summary, \method abstracts the domain knowledge and user behavior expectations in a form that an LLM agent can consume. It acts as shared documentation for both the orchestrator and the TCA. In the next subsection, we describe how exactly this document is reorganized and utilized in the system prompts.

\begin{figure*}[t]
 \centering
 \includegraphics[width=0.98\textwidth]{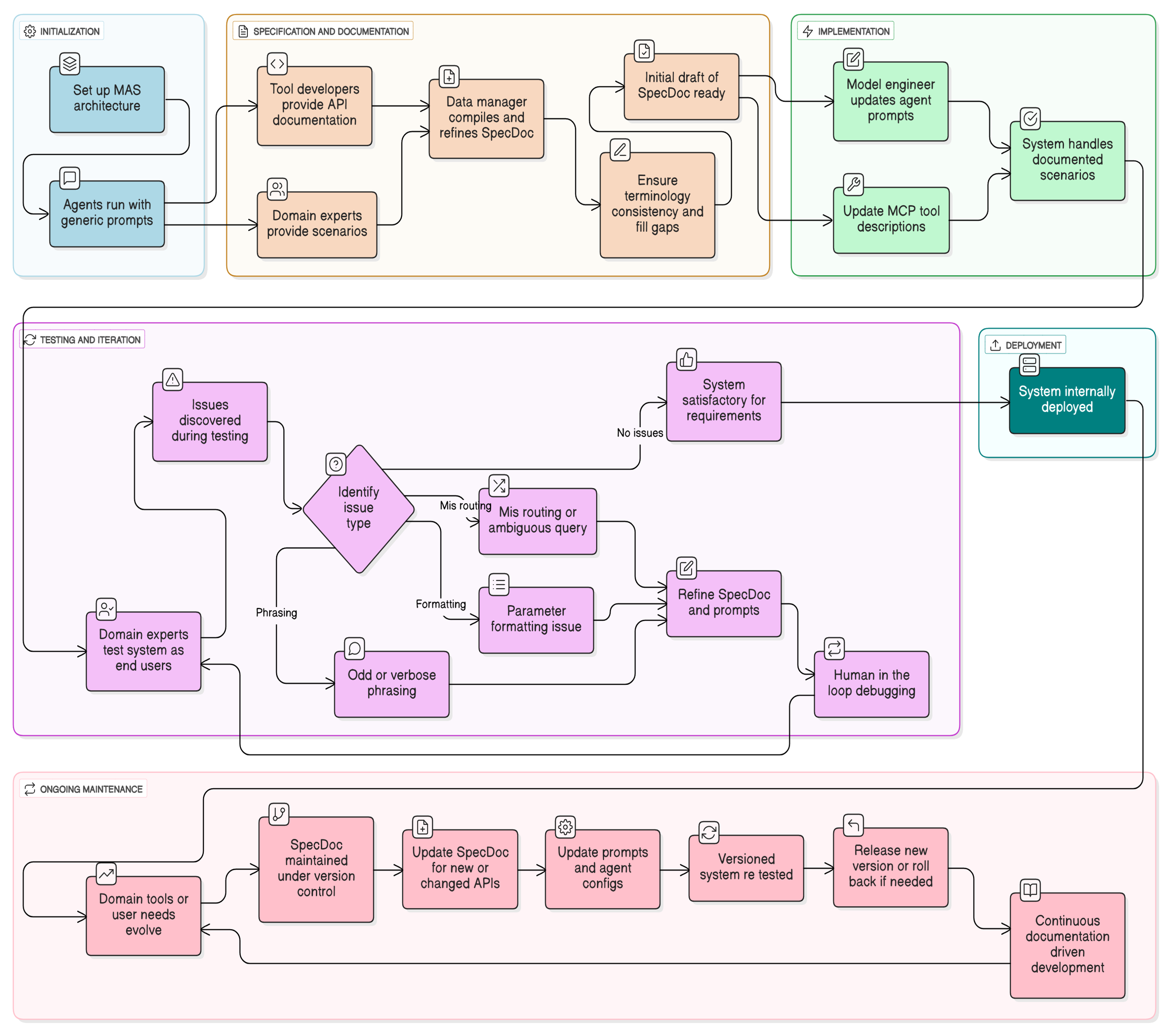}
 \caption{Development and maintenance flow of \framework on a private-domain multi-agent system. \method is created from domain tool definitions and expert scenarios and fed into the instructions of each agent and API docstrings, and iteratively refined after evaluation of domain experts. Updates to tools or domain knowledge are handled by editing \method and updating agent prompts accordingly, with version control.}
 \label{fig:workflow}
 \end{figure*}

\subsection{Incorporating \method into Agents' Prompts and Docstring}
\label{sec:spec2agent}

Once \method for the domain tools are prepared, the next step is to inject the relevant parts of it into $S$. In our design, this involves two steps: the TCA’s prompt and the documentation strings (docstrings) for the tool APIs, which the TCA will have access to when deciding on and executing a tool call. Additionally, a portion of the specification knowledge might be embedded into the orchestrator’s prompt to aid routing decisions.
When the orchestrator decides a query that should be handled by the TCA, it will pass along the conversation history to the agent. The TCA’s system prompt contains general instructions on identifying the intention of user queries, domain-specific jargons and  the detailed guidelines for choosing the appropriate tool, extracted from \method, e.g., the \textit{``input arguments and related expressions”} and \textit{``when to use which tool”} parts. For example, the TCA prompt might look like:

\begin{quote}\small
\textit{ “You are a tool-calling agent for domain $D$. You have access to the following tools: \texttt{[Tool A, Tool B, ...]}.} 

\textit{\textbf{Your responsibilities:}}

\textit{- Understand the user’s request in context,}

\textit{- Identify which tool (or tools) can address it,}

\textit{- Convert the user request into the appropriate API call(s) with correct parameters,}

\textit{- Execute the call(s) and gather results,}

\textit{- Formulate a helpful answer using the results (or an error message if the tool fails).}

\textit{\textbf{Guidelines:}}

\textit{- If the user uses domain-specific jargon or nicknames, normalize them according to the documentation.}

\textit{- If multiple tools are needed, you may call them in sequence.}

\textit{Only call tools relevant to the query; do not call tools for chit-chat or general questions. ... (and more)}

\textit{\textbf{Tool Specifications:}}

\texttt{[Tool A – Purpose: ...; Inputs: ... (with synonyms: ...); Outputs: ...; etc.]}

\texttt{[Tool B – Purpose: ...; ...]}

\textit{Use the above specs to decide your actions.”}
 \end{quote}

When TCA actually chooses a tool and calls it via LLM-generated argument json, we assume there is an underlying mechanism that will execute the call.
In our case, for a private deployment, we assume an actual MCP client integrated: this client has docstrings or schema for each API reflecting the specification. We thus ensure that the LLM’s action of calling a tool is guided by the same information. Essentially, each API endpoint in the MCP server has a self-documentation that is derived from SpecDoc. The process of placing information involves carefully curating which parts of the document each component sees, but usually the best case is placing all relevant information that is not used for the TCA prompt, since docstring is not usually counted as initial agent prompt and only activated when the tool is chosen. 
For instance, the MCP docstring might describe the chosen tool as:

\begin{quote}\small
GetMachineStatus(machine\_id: str) -> StatusInfo: Returns the operational status of a machine given its ID

- \textit{machine\_id (str)} is usually an alphanumeric expression with three alphabets and two digits, accompanying user expressions such as `equipment', `device', etc.

- \textit{StatusInfo} includes fields \{\textit{status}, \textit{last\_update\_time}\} where \textit{last\_update\_time} is a timestamp for the last machine update, in format YYYYMMDDSS, and \textit{status (str)} denotes the comment written at \textit{last\_update\_time} by the engineer who handled the machine.

- Use this when the user asks about a machine's status or whether a machine is running/stopped (e.g., ``Is XG242 up?''). The machine may be referred to by only the number if the user profile contains any information about the characteristics of the machine (e.g., ``XG'' denotes a specific engineer site).
\end{quote}

We found it useful to include a few canonical examples in the prompt as well, illustrating user queries and the corresponding tool call, as in demonstration approaches \cite{min2022rethinking}. These can be taken from the scenarios provided by domain experts or when the system is deployed, extracted from the real user queries. Including 1–2 shot examples in the prompt can significantly boost the reliability of the agent’s understanding, effectively serving as inline tutoring.

The orchestrator’s prompt might also benefit from \method, but a lighter touch is required because the orchestrator only needs to classify intent and route it to agents. Orchestrator can extract from the spec document a list of keywords or patterns that indicate domain-related queries, for example: \textit{``If the user query mentions any machine IDs, equipment numbers, failure rate, or other manufacturing-specific terms in [list provided], route to the tool-calling agent. If the query is a general greeting or outside domain, route to general chat agent.”} These cues can be extracted from the jargon lists of the spec document. 
Essentially, we give the orchestrator a high-level domain vocabulary to help it identify relevant queries. 
Through this careful context engineering, we integrate the domain knowledge directly into the behavior of the orchestrator and the TCA. In doing so, we achieve what previously might require thousands of training examples: the orchestrator knows the boundaries of when to invoke the TCA, and the TCA knows how to transform user language into correct API calls. The difference is all the knowledge is explicit and maintained in one place (\method) rather than spread implicitly in a dialogue-trained model.

% \usepackage{graphicx}
\begin{table*}[]
\makegapedcells
\centering
\caption{Comparison of two in-house approaches of LLM-based private-domain tool-calling systems. Note that PM of `Data building' indicates person-month, and `LLM used' are all open-weighted models (the final selection was done based on the performance, within the limited parameter size). 
}
\label{tab:inhouse}
\begin{tabularx}{\textwidth}{|l|X|X|}
\hline
\textbf{} &
  \multicolumn{1}{c|}{\textbf{Single-agent, A fine-tuned LLM}} &
  \multicolumn{1}{c|}{\textbf{Multi-agent, No LLM post-training}} \\ \hline
\textbf{Coverage} &
  8 APIs, 2 subdomains &
  14 APIs, 6 subdomains \\ \hline
\textbf{Architecture} &
  A single PEFTed LLM with a template-style prompt &
  A single vanilla LLM with multiple prompts  % {\footnotesize (2$\sim$3 used, fit for each purpose, using a single LLM is also capable)} 
  \\ \hline
  \textbf{Prompts} &
A partially variant prompt with the system and tool-argument description (which was also used in tuning) &
Prompts for the orchestrator, TCA, and GCA, along with MCP docstrings  % {\footnotesize (2$\sim$3 used, fit for each purpose, using a single LLM is also capable)} 
  \\ \hline
\textbf{Supports} &
  \begin{tabular}[c]{@{}l@{}} Daily conversation and tool-calling \\ (tuned with synthetic dialogues)\end{tabular} &
  \begin{tabular}[c]{@{}l@{}} Daily conversation and tool-calling \\(Orchestrator + 2 sub-agents)\end{tabular}\\ \hline
\textbf{Resource used} &
  Use high resource in tuning, less resource in service &
  Use adequate resource in service, and none in tuning \\ \hline
\textbf{LLM used} &
  Up to 30B models for PEFT and inference &
  Up to 120B models only for inference \\ \hline
\textbf{Dataset} &
  $\sim$1.3K tool-calling dialogues with 1$\sim$5 turns &
  \method for 14 APIs within 6 subdomains \\ \hline
\textbf{Data building} &
  Built over 8 months by two developers and three domain API engineers (40PM) &
  Built over 3 months by two developers, three domain API engineers, and five non-developer domain experts (30PM) \\ \hline
\textbf{API args} &
  3$\sim$10 argument types, with high portion of argument overlap between APIs &
  3$\sim$20 argument types, with high portion of argument overlap between APIs \\ \hline
\textbf{Evaluation} &
  \begin{tabular}[c]{@{}l@{}} Single-turn, 480 queries \\  {\footnotesize (endpoint acc 87.9\%, slot-filling acc 77.5\%)}\end{tabular} &
  \begin{tabular}[c]{@{}l@{}} Single-turn, 540 queries \\  {\footnotesize (endpoint acc 97.0\%, slot-filling acc 82.6\%)}\end{tabular} \\ \hline
\textbf{Flexibility} &
  Less flexible due to limited scenario support, so the behavior is more close to inductive one, and some entity names can bring inductive bias &
  Highly flexible since the conversation is performed based on definition and instruction, so the behavior is more close to deductive, one and the system has more comprehensive analysis on unseen entity names \\ \hline
  \textbf{Maintenance cost} &
High (Dataset rework required for each tool and convention update) &
Relatively low (\method update followed by agent prompt and docstring update) \\ \hline
\end{tabularx}
\end{table*}

\subsection{Development and Maintenance}

Figure~\ref{fig:workflow} illustrates the overall workflow of developing and maintaining our behavior-modeled multi-agent system $S$. The process begins with the \textit{system developers} setting up the basic multi-agent architecture (Orchestrator, TCA, GCA, plus the MCP integration). 
\smallskip\\
(1) Initially, these agents run with generic prompts. Next, the \textit{tool providers} and \textit{domain experts} contribute by providing documentation for the tools and usage scenarios. This often starts as a collection of use cases: e.g. \textit{``User asks for an equipment with the highest failure rate $\rightarrow$ system should call EquipmentAPI $\rightarrow$ return the equipment with the most frequent stops in recent two weeks’’}. Domain experts essentially enumerate what they expect the AI assistant to handle. Tool providers construct the technical details of the APIs – their formal specifications, valid values, etc.
\smallskip\\
(2) The data builders of \textit{system developers} group then compiles this information into \method as described. They ensure consistency in terminology and fill in any gaps – such as adding likely user phrasings if the domain experts did not specify them explicitly, based on the discussion on how users speak in $D$. This writing yields an initial draft of \method.
\smallskip\\
(3) Now the model engineer of \textit{system developers} group updates the agent prompts and the MCP tool descriptions accordingly based on \method (as discussed in Sec~\ref{sec:spec2agent}) so that $S$ can handle the documented scenarios. 
\smallskip\\
(4) The next step is iterative inspection: the \textit{domain experts} evaluate the system $S$ by interacting with it as \textit{end-users}, hopefully via available UI. They check if the orchestrator correctly routes queries, if the TCA manages to call the correct APIs with appropriate arguments, and if the GCA covers the rest. Inevitably, this testing may reveal some typical issues: the orchestrator might mis-route an ambiguous query; TCA might format a parameter incorrectly; or the responses might be correct but phrased oddly since the LLM might use too verbose a style or not enough detail. The issues are addressed by iteratively refining \method and system prompts. For example, if a mis-routing occurred due to the absence of a certain phrase, we can add it to the orchestrator’s instruction. If a parameter formatting issue occurs, we might update \method to emphasize the expected format or add an example of that case. Essentially, this human-in-the-loop process is analogous to how the data manager would have debugged in the traditional pipeline by adjusting rules.
\smallskip\\
Once the execution of $S$ is satisfactory regarding the initial requirements, we consider the system ``deployed” internally. However, we anticipate that tools will be added, removed, or updated in domain $D$ over time. Also, \textit{end users} might start using the assistant in novel ways not anticipated in the original specification. 
Thus, we recommend maintaining \method under version control. Whenever a change is necessitated — say a new group of APIs is introduced — the data manager updates the document (adding a new section or modifying an existing section), also highlighting the differences. The orchestrator and the TCA prompts are then updated to include the new/modified information. 
This versioning approach treats the combination of (\method, prompts, agents) configurations as a ``model” that evolves, instead of tuning the weights of a neural network.
In contrast to de facto where retraining should be conducted for changes, here the ``knowledge base” of the agent is the living document which can be edited and uploaded to the agent instantly.

\begin{table*}%[!b]
\makegapedcells
  \small
  \centering
  \caption{Comparative analysis of system-building approaches. 
  }
  \label{tab:comparison}
  \begin{tabularx}{\textwidth}{|>{\raggedright\arraybackslash}p{3.6cm}|*{7}{>{\raggedright\arraybackslash}X|}}
    \hline
    %& \textbf{Methods}
    & \textbf{System Complexity ($\downarrow$)}
    & \textbf{Data Required ($\downarrow$)}
    & \textbf{Compute Resources ($\downarrow$)}
    & \textbf{Manual Labor ($\downarrow$)}
    & \textbf{Flexibility ($\uparrow$)}
    & \textbf{Forgetting Risk ($\downarrow$)} \\
    \hline
    \textbf{Classical Modular pipeline}
      %& Rule-based or ML models for NLU, DST, Policy, NLG
      & Medium %(many components, integration needed)
      & Moderate %(some annotated data for NLU, etc.)
      & Low %(models are small or rules)
      & High %(hand-crafting rules, maintaining pipeline code)
      & Low %(hard to add new intent or slot without code changes)
      & N/A %(no large model to forget, but might have inconsistent rules) 
      \\
    \hline
    \textbf{LLM Fine-tune, Single-agent}
      %& One large model fine-tuned on domain dialogues
      & Low/Med %(single model, simpler pipeline)
      & Very High %(dialogue data covering tool use scenarios)
      & High %(fine-tuning LLM and running it)
      & High %(creating synthetic dialogues or annotation for fine-tuning)
      & Low/Med %(model may need retraining for new tools or significant changes)
      & High %(catastrophic forgetting of general capabilities when fine-tuned on narrow data \cite{zhang2023forgetting}) 
      \\
    \hline
    \textbf{LLM Fine-tune, Multi-agent}
      %& Multiple LLMs, each fine-tuned for role (or jointly via multi-task)
      & High %(multiple models + coordination logic)
      & Very High %(data needed for each agent's role)
      & Very High %(fine-tuning and running multiple LLMs)
      & Very High %(same as single-agent, multiplied by number of agents)
      & Low %(even more retraining needed for changes)
      & High %(each agent fine-tuned could overfit to domain and lose adaptability) 
      \\
    \hline
    \textbf{LLM + \method (Ours)}
      %& Pre-trained LLMs orchestrated with prompts and documentation
      & High %(multiple agents, but no model training, uses prompting)
      & Low %(no training data; requires comprehensive documentation)
      & Medium %(running a few LLMs inference; no fine-tune phase)
      & Medium %(writing and updating spec document, prompt tuning)
      & High %(spec doc and prompts can be edited for updates; quick iteration)
      & Low %(base LLM remains unchanged, general ability preserved; domain info provided on the fly) 
      \\
    \hline
  \end{tabularx}
\end{table*}

\subsection{Training-Free Building of Private Domain Multi-Agent Systems}

The primary use case of our framework is to deploy a functional multi-agent assistant without any model training, purely by providing the necessary domain knowledge through specifications. This is especially valuable for organizations that cannot or prefer not to fine-tune large models on their data, due to data sensitivity, lack of ML infrastructure, or concerns about model drift. Using our approach, an enterprise can stand up a conversational agent that interfaces with internal APIs in a matter of days, by writing \method and prompts, rather than months of data collection and iterative training. This lowers the barrier to adoption of AI assistants behind company firewalls. Moreover, the resulting system can be run on-premise (using open-sourced LLMs and frameworks if needed), satisfying strict privacy requirements. 

Here, we provide a comparison of our in-house implementation approaches of the system (Table~\ref{tab:inhouse}). We built an MAS for manufacturing industry users within a private domain, while our initial attempt had been a simple single agent system based on LLM fine-tuning. We found that the training-free MAS achieves a higher success rate on the predefined scenarios even for more complicated scenarios (8 APIs to 14 APIs) with more limited person-month (40PM to 30PM), accompanied with thorough \method refinement. Though the execution of both approaches were not conducted simultaneously, we checked that the performance of training-free MAS was on par with what a fine-tuned model might have achieved in those scenarios, and in more a flexible and efficient way. In practice, as \method includes detailed user behavior, the LLM follows it quite strictly, likely because providing the model with well-structured knowledge in its context steers to use that knowledge rather than guessing.

Our comparison has several limitations though -- first, the implementation of both systems were not conducted simultaneously, so the developers or workers' empirical knowledge in modeling the behavior of the system and users might have been affected by the experience of dataset construction performed at the first place. Second, due to the intrinsic difference of PEFT-based and inference-only systems, our model usage for both approaches differed in their parameter size and capabilities. This also concerns the different timeline of both approaches (about 8 months) -- where considerable improvements of open-weighted LLMs took place.
Lastly, throughout the development, the update of API was inevitable for the user-centric execution of the system -- we did not perform the comparison with the same API set and specification, and instead reported the different context of each approach. Due to the differentiation of API set and specification, the pool and range of evaluation was also modified, which did not reach in fully fair comparison. 

All the limitations of our comparison owes to the product-centered nature of the AI development within the enterprise. Though we acknowledge this -- and that our empirical result is not based on lab condition, our preliminary results may serve as a practical reference for in-house building of training-free multi-agent systems.

\subsection{Comparison with Traditional Approaches}

For practitioners, we illustrate a conceptual comparison between our \method + MAS approach and prior methods for building conversational systems: (a) the classical modular dialogue system (with separate NLP components, not LLM-based), (b) an end-to-end single-agent LLM fine-tuned on domain data, and (c) a multi-agent LLM system where each agent is fine-tuned on domain data (Table~\ref{tab:comparison}). We compare across several criteria that matters in usual domain-specific systems, to display that our approach takes the advantage of high flexibility with lower labor, at the same time lessening the risk of forgetting general capability.
We eliminate the need for ongoing model retraining, instead shifting to a maintenance-of-documentation paradigm. Even for tasks with complex decision logic that an LLM cannot infer from documentation alone, we argue that with adequate exploitation of its tool use and reasoning capabilities, e.g. using planning algorithms in the loop if necessary, one can cover a wide range of private-domain needs without training new models.

Looking ahead, we expect that private domain vertical AI systems will increasingly adopt this kind of methodology. Training and deploying a bespoke LLM for every enterprise is not feasible nor economic; rather, leveraging powerful foundation models with in-context alignment, through specification documents, tools, and multi-agent decomposition, is a more sustainable path. It keeps the cutting-edge LLMs' general knowledge intact while allowing deep customization to proprietary data and tools. Our framework demonstrates one way to achieve that for conversational agents that must perform actions in a private domain.

\section{Discussion}

Though we've demonstrated our implementation detail and the use case with preliminary results, some challenges still limit the in-house adoption of our framework and method.

\subsection{Challenges}

First, the current process of manually writing up the \method placing the specification into prompts and docstrings is labor-intensive and error-prone. We envision that in the future, one could automate this by creating a draft documentation from the very first version of APIs, analyzing user queries based on the given tools to finalize a structured document, and parsing them to adequate agent prompts and docstrings.

Another hurdle of the proposed approach is that it relies highly on the long context which might be a constraint if available models are small or if \method contains too long body of docstrings. Thus, in practice, one can mitigate this by splitting tools into groups and only loading the relevant group’s specification, possibly splitting TCA into sub-agents as well. In our manufacturing example, if tools are grouped by their function (e.g. production data vs. maintenance logs vs. staffing info), the orchestrator could first pick a group, then TCA loads only that group’s specs. This engineering optimization can be further implemented by adopting recent documenting techniques such as skills\footnote{\url{https://platform.claude.com/docs/en/agents-and-tools/agent-skills/overview}}. 

Other critical points include the human conflict which is also widely observed in human-in-the-loop data construction processes, especially within the domain expert group. Since they are not familiar with executing the domain-specific tools in natural language manner, some conflicts become apparent when they argue with each other about the convention of jargon usage and the desired action of the system. This also leads the conflict to the issue of proper evaluation, which is already a challenge of multi-turn, multi-agent systems. Beyond simply testing single-endpoint queries, further evaluations should incorporate the comprehensive multi tool-calling cases and failure-trial circumstances.

\subsection{Lessons Learned}

Despite challenges addressed, we want to share some lessons learned from our development and deployment experience:

\begin{itemize}
    \item \textbf{Sustainability:} Well-built \method can be utilized sustainably regardless of the advance of LLMs and system architectures if it includes the domain-specific conventions that are unchanged despite tool updates, not covered within the scope of generic knowledge.
    \item \textbf{Human labor:} Although public LLMs' capabilities showed boost in specialized domains, building agentic system within enterprise-level applications (especially those with compliance and security policies) still requires human labor and manual inspection for visible functioning due to undocumented conventions.
    \item \textbf{Moderation:} As the importance of human labor and inspection is still significant, the moderation phase for practitioners (e.g., between domain experts, between domain and system) and the language translation between stakeholders are important as well for the feasible implementation of the system.
    \item \textbf{Failures help:} While the deductive \framework seems efficient and flexible compared to the traditional data synthesis and tuning approach, writing up the working version of \method would have not been possible for the failures and errors experienced in the previous dataset building.
\end{itemize}

The technology develops rapidly and the industry is already familiar with the concept of what we've demonstrated so far, such as aforesaid skills, the context engineering which is the advanced and comprehensive form of prompting, and the recent harness engineering that controls malfunction of LLMs in multi-agent environment. However, we believe that common tools still do not cover undocumented conventions, and as the portion of automated system building increases, the importance of expert intervention will be more valuable, especially and ironically in most enterprise settings where the proprietary information becomes the power.

\section{Conclusion}

We have introduced a framework \framework for building private-domain conversational agents that forgoes training in favor of explicit user and system behavior modeling through documentation. 
Also, we presented a method \method to align LLM-based multi-agent systems to private domains using structured documentation instead of model fine-tuning that requires large-scale data generation -- which draws on the strengths of both classical software engineering and modern LLM abilities. 
We believe this approach offers a sustainable path for vertical AI systems: it keeps human experts in the loop via documentation and avoids repeatedly re-training models whenever requirements change. It also fosters transparency – \method can be reviewed by stakeholders – including those concerned with governance or compliance – to verify the system’s intended behavior.

Our approach has several limitations --
Firstly, our work is so far empirical; we did not present quantitative benchmarks or a deployed large-scale user study. In part, this is because private domains by nature lack an evaluation scheme that can be disclosed, and the trials are in-house and practical. Besides, while our approach reduces the risk of forgetting since we don’t fine-tune, it doesn’t eliminate all failure modes of LLM-based agents. The LLM could still hallucinate an API call that doesn’t exist if the document is incomplete or if it misinterprets. It could output an answer that is factually incorrect even if the tool result was correct. We mitigated these by demonstration examples and carefully written prompts, but evaluation in a real setting is needed to identify if any systematic issues remain. 

Going forward, as LLM technology improves, such training-free alignment might become even more viable. With models that can ingest larger contexts or plug in tools more seamlessly, the role of humans might shift to mostly curating and updating knowledge. Our framework is an early step in that direction, applied to the multi-agent conversational setting. We hope this inspires more research on how to best combine human knowledge encoding with LLM reasoning for building robust, domain-specific AI agents.

\section*{Data Availability Statement}

We are not disclosing the data used due to the proprietary constraints. However, the detailed composition of document templates and how they should be placed into code scripts are presented in the manuscript with a virtual API as an example. We believe practitioners building their own enterprise-level agentic wrappers would directly benefit from our demonstrations given that their own tools are prepared. We plan to construct and disclose such tool-to-context transformation pipeline for future human-in-the-loop automation.

\begin{acks}
We thank Soonmo Kwon, Junwoo Song, Hodong Lee, Sungjoo Suh, Jun Haeng Lee, and Team ONIX of Data Analysis Platform (A-FAB) for grateful assistance and comments in building our system.
\end{acks}

\bibliographystyle{ACM-Reference-Format}
\bibliography{sample-base}


\end{document}